\documentclass{pasj00}
\draft

\newcommand{\feoh}{\textrm{[Fe/H]}}
\newcommand{\cfe}{\textrm{[C/Fe]}}
\newcommand{\nucm}[2]{\ensuremath{{}^{#1}{\rm #2}}}
\newcommand{\sun}{\ensuremath{\odot}}
\newcommand{\msun}{\ensuremath{M_{\odot}}}
\newcommand{\loge}{\ensuremath{\log \epsilon}}
\newcommand{\teff}{\ensuremath{T_{\textrm {eff}}}}
\newcommand{\nobj}{1212}

\begin{document}
\SetRunningHead{Suda et al.}{The Stellar Abundances for Galactic Archaeology (SAGA) Database}
\Received{2008/04/08}
\Accepted{2008/05/30}

\title{The Stellar Abundances for Galactic Archeology (SAGA) Database
-- Compilation of the Characteristics of Known Extremely Metal-Poor Stars}

\author{Takuma \textsc{Suda},\altaffilmark{1}
        Yutaka \textsc{Katsuta},\altaffilmark{1}
		Shimako \textsc{Yamada},\altaffilmark{1}
		Tamon \textsc{Suwa},\altaffilmark{2}
		Chikako \textsc{Ishizuka},\altaffilmark{1}
		Yutaka \textsc{Komiya},\altaffilmark{3}
		Kazuo \textsc{Sorai},\altaffilmark{1}
		Masayuki \textsc{Aikawa},\altaffilmark{4}
		Masayuki Y. \textsc{Fujimoto},\altaffilmark{1}
}

\altaffiltext{1}{Department of Cosmosciences, Hokkaido University, Kita 10 Nishi 8, Kita-ku, Sapporo 060-0810, Japan}
\altaffiltext{2}{Center for Computational Sciences, University of Tsukuba, Ten-nodai, 1-1-1 Tsukuba, Ibaraki 305-8577, Japan}
\altaffiltext{3}{Department of Astronomy, Graduate School of Science, Tohoku University, Sendai 980-8578, Japan}
\altaffiltext{4}{Hokkaido University OpenCourseWare, Hokkaido University, Kita 11 Nishi 5, Kita-ku, Sapporo 060-0811, Japan}

\KeyWords{astronomical data bases: miscellaneous${}_1$ --- stars: abundances${}_2$ --- stars: evolution${}_3$}

\maketitle

\begin{abstract}
We describe the construction of a database of extremely metal-poor
(EMP) stars in the Galaxy. Our database contains detailed elemental
abundances, reported equivalent widths, atmospheric parameters,
photometry, and binarity status, compiled from papers in the
literature that report studies of EMP halo stars with $\feoh \le -2.5$.
The compilation procedures for this database have been designed to
assemble the data effectively from electronic tables available from
online journals. We have also developed a data retrieval system that
enables data searches by various criteria and illustrations to
explore relationships between stored variables. Currently, our
sample includes 1212 unique stars (many of which are studied by
more than one group) with more than 15000 individual reported
elemental abundances, covering the relevant papers published by
December 2007. We discuss the global characteristics of the
present database, as revealed by the EMP stars observed to date.
For stars with $\feoh \le -2.5$, the number of giants with
reported abundances is larger than that of dwarfs by a factor
of two. The fraction of carbon-rich stars (among the sample for
which the carbon abundance is reported) amount to $\sim$ 30 \%
for $\feoh \le -2.5$. We find that known binaries exhibit different
distributions of orbital period, according to whether they are
giants or dwarfs, and also as a function of metallicity,
although the total sample of such stars is still quite small.

\end{abstract}

\section{Introduction}

Extremely metal-poor (hereafter EMP, defined by $\feoh \leq -2.5$ in this paper)
stars in the Galaxy carry information about the physical conditions in the early
epochs when they were born, and are also unique probes of the production of
elements by the first generation stars when the Universe emerged from the
so-called dark ages. Analysis of their kinematics also provides direct
information on the early stages of galaxy formation (e.g., \cite{Carollo2007}).
The chemical compositions of these stars also impose constraints on the
nucleosynthesis pathways involved with the formation of the elements throughout
the early history of the Galaxy.

It is a long standing problem whether one can identify the low-mass survivors of
the first-generation (Population III) stars, those objects born from primordial
clouds containing no elements heavier than lithium. If such stars did form it
remains possible that they could be found among the EMP stars, since there are
processes (such as binary mass-transfer and/or the accretion of interstellar gas
polluted by later generation stars) that could effectively ``disguise'' their
true nature by making them appear more metal-rich at present.

Thanks to the recent large-scale searches for candidate Very Metal-Poor stars
(hereafter VMP, defined by $\feoh \leq -2.0$ according to the nomenclature
of Beers \& Christlieb 2005),
in particular
by the HK survey \citep{Beers1985, Beers1992} and by the Hamburg/ESO survey
\citep{Christlieb2008}, the number of known VMP stars has increased
dramatically since the 1990s. Approximately $\sim 1200$ and $\sim 1500$ stars
have been identified as VMP to date, on the basis of medium-resolution
spectroscopic follow-up of the $\sim 6000$ and $\sim 3600$ candidate
in HK-I and HES survey, respectively \citep{Beers2005b}.
This number is likely to expand quickly, as additional VMP
stars are identified from ongoing efforts such as the Sloan Digital Sky Survey
(in particular from SEGUE: Sloan Extension for Galactic Understanding and
Exploration, see http://www.sdss.org). Furthermore, high-resolution
spectroscopic observations with 8m-class telescopes such as SUBARU, the VLT, and
the KECK telescopes are already beginning to elucidate the detailed abundance
patterns of VMP stars.

The abundance analyses of EMP stars provide useful information on Galactic
chemical evolution by comparison of their abundance patterns with those of more
metal-rich stars with $\feoh \gtrsim -2.5$, including the globular cluster
stars. At present, only three stars are known with metallicities well below
$\feoh = -4$ (all of which have high-resolution abundance analyses available),
while more than 100 stars are known with $\feoh < -3$, roughly half of which
have detailed abundance analyses at present. A salient feature of EMP stars is
the sharp cut-off below $\feoh \sim -3.5$ in the metallicity distribution
function. Other important features of EMP stars are the large fraction of
carbon-enhanced stars that are known to exist among them, especially below
$\feoh \sim -2.5$
\citep{Rossi1999}, as well as the large scatter in the abundances of
neutron-capture elements \citep{Gilroy1988, McWilliam1995b, Ryan1996,
Francois2007}. The lighter elements, such as CNO, as well as the {\it p}- and
$\alpha$-capture elements and {\it s}-process elements, are used as tools to
explore nucleosynthesis from H- and He-burning resulting from binary mass
transfer affected by the evolution of low- and intermediate-mass AGB stars
\citep{Suda2004, Lucatello2006,Komiya2007}. For heavier elements, the abundance
patterns of individual EMP stars provide crucial information on the
r-process elements produced (presumably) by individual supernova events
\citep{Truran1981,Mathews1990}.
Such stars are also used as cosmo-chronometers for placing lower limits on the
age of the Universe, based on determinations of the abundances of uranium and
thorium \citep{Sneden1996,Wanajo2002}. A handful of stars that exhibit large
enhancements of the r-process elements \citep{Sneden1994,Hill2002, Frebel2007a} have
drawn the interest of researchers concerned with nucleosynthesis processes in
massive EMP stars and the chemical evolution of the Galaxy. The
determination of the isotopic abundances of
\nucm{6}{Li} and
\nucm{7}{Li} by high-resolution spectroscopy \citep{Smith1993,Hobbs1994,Asplund2006}
also impacts observational constraints on Big Bang nucleosynthesis and the
astrophysical origins of these elements.

In order to promote studies such as those described above, and to make them more
useful in aggregate (e.g., for statistical studies), it is desirable to
construct a modern database of the elemental abundances (and other related
properties) of metal-poor stars in our Galaxy. Although the available data on
the abundances and properties of EMP stars has been greatly increasing in
the past decades, thanks to the many high-resolution spectroscopic studies
that have been conducted 
\citep{Gilroy1988,Ryan1991,McWilliam1995b,Ryan1996,Fulbright2000,Preston2000,Burris2000,
Mishenina2001,Aoki2002b,Cohen2002,Carretta2002,Johnson2002,Nissen2002,Cayrel2004,Honda2004b,
Cohen2004,Simmerer2004,Spite2005,Barklem2005,Jonsell2005,Aoki2005,GarciaPerez2006a,Cohen2006,
Aoki2007b,Francois2007}, there are no present databases that make these data readily
available to astronomers and other researchers in order to conduct their own
studies. Generally, it is quite difficult (in particular for the non-specialist)
to collect the relevant quantities from the widely scattered literature. In
part, this is because the data are presented in individual papers using wide
varieties of formats, such as text, tables, and figures. Therefore the
compilation of pertinent information requires a great deal of human resources
for individual investigators, who would benefit greatly from a more automated
compilation.

To develop a more effective set of tools for compilation of data for
EMP stars, we have adopted a similar set of methodology for data compilation
developed by the Japanese nuclear data group\footnote{The Stellar Abundances for Galactic Archeology database, SAGA.
The database will be available at http://saga.sci.hokudai.ac.jp/.}
\citep{Suda2006b}.
We adopt the strategy of Hokkaido University Nuclear Reaction Data Center
(JCPRG), which has developed tools for compilation via the internet that
alleviate much of the human resources required if the data were input manually
from the literature \citep{Otuka2002}. We differ from the JCPRG approach in that we have
adopted a relational database management system for data storage ({\it MySQL}),
rather than a text-based master database. We have also adopted their methods of
utilizing the database through the internet by developing the tools to retrieve
data and draw summary graphs \citep{Nouri2002,Otuka2005,Pritychenko2006}.

In this paper we describe the structure of the SAGA database, and present some
results based on simple queries of the existing system. Our database enables
queries of quantities such as the elemental abundances, photometry, atmospheric
parameters, binarity, and position in the Galaxy, and the relationships between
them. Thereby, we can begin to consider the characteristics of EMP stars in a
statistical sense, and better draw global views of the nature of EMP stars in
the Galaxy.

The paper is organized as follows. In \S 2 we describe the
compilation and retrieval system for our database. In \S 3 we elaborate on the
global characteristics of EMP stars in our sample.
In \S 4 we present a brief summary.

\section{Overview of the Database and the Retrieval System}

\subsection{Characteristics of the SAGA database}

The SAGA database assembles available data for extremely metal-poor stars in the
Galaxy from the recent literature, and makes them available for observational
and theoretical studies. At present, we have collected papers so as to cover a
complete sample of metal-poor stars in the Galaxy with $\feoh \leq -2.5$,
excluding stars in metal-poor Galactic globular clusters.
At the same time, we compile the reference stars and stars with $\feoh > -2.5$
listed in the same literature along with EMP stars .
This metallicity range
corresponds theoretically, for evolved stars, to the occurrence of
hydrogen-mixing into the helium convection zone during helium shell flash and
the helium-flash driven deep-mixing episodes in the helium burning region
\citep{Fujimoto2000}, which plays a critical role in understanding the
characteristics of EMP stars.

Thus far we have compiled data for \nobj\ stars (based on 2243 independently
reported measurements of stars, 1031 of which are included in multiple papers)
from 121 papers published through December, 2007.
Among them, the number of stars with $\feoh \leq -2.5$ amounts to 392, although
it varies with the adoption of derived iron abundance from multiple papers.
The stellar parameters and
abundance data we have collected are based mostly on high-dispersion
spectroscopy (typically $R \gtrsim 40000$), although some data obtained with
lower dispersions are also included. We intensively gathered data from the
latest papers, and currently trace back to 1995. We believe we have complete
coverage for papers published from 2000 onward, except for a few redundant
examples for which electronic data are only available on the Web. 

The quantities compiled in the database are summarized in Table~\ref{tab:quantity}.
The data cover the stellar parameters and photometric data, as well
as information gleaned from papers that included a log of observations. During
the course of compilation, a unique entry number is assigned to each individual
paper. Bibliographic data are identified by the title, authors, and reference
code of the selected paper. The compiled data are stored as a table, following
the example shown in Table~\ref{tab:quantity}. Positions of the stars are
obtained from the literature, the SIMBAD database, or the VizieR catalogues.
Galactic coordinates are calculated, if required, and are also placed in the
database. The data from the observing logs contain basic information concerning
the observational set up that was employed, and may be used to obtain a quick
review of individual stars. The records of (heliocentric) radial velocities are
stored along with their dates of observation, which should prove useful for
checking the status of monitoring programs for binarity, as well as for use in
studies of the space motions of the stars, when combined with proper motions
(some of which are presently available, and others that should come available in
the future).
We have compiled the adopted solar abundance in the literature that enables us
to evaluate the effect of using different solar abundances. If we remove
this effect, we can obtain more homogenous data for relative abundances
independent on the adopted solar abundances.
The abundances of individual elements in our collected sample
constitute more than 17,000 records for species from lithium to uranium.

\begin{figure}
  \begin{center}
    \FigureFile(150mm,){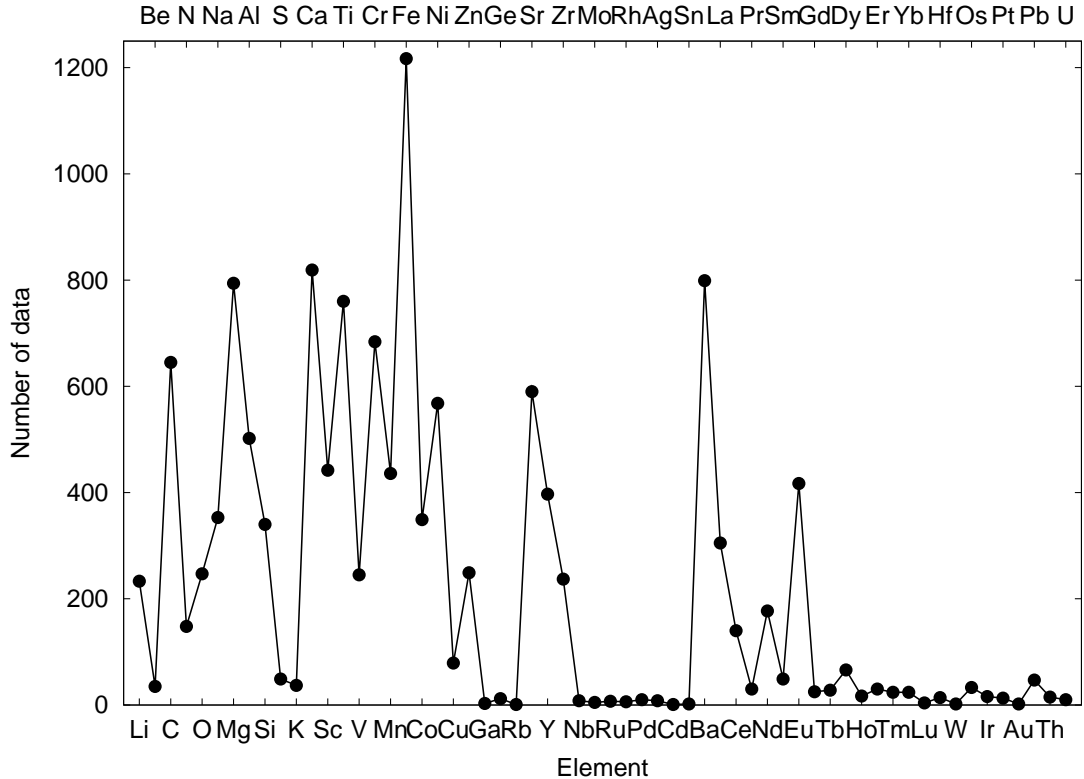}
  \end{center}
  \caption{The number of compiled data as a function of chemical species.
  }\label{fig:element}
\end{figure}

Figure~\ref{fig:element} shows the number of stars that have available abundance
data for each element, which clearly varies greatly depending on the species
considered. We distinguish abundance information based on fits to synthetic
spectra of atomic and molecule lines from that based on equivalent width
measurements. These are stored as separate records. The data for reported
isotopic abundances are also recorded separately in our database, although we
have limited this information (for now) to the ratios $^{12}{\rm C}/ ^{13}{\rm
C}$ and $^{6}{\rm Li}/ ^{7}{\rm Li}$. The stellar atmospheric parameters used to
derive the spectroscopic abundances are available in most papers, and are
compiled in the database. Photometric information (magnitudes and color indices), 
if available, is compiled from literature. These are also useful for
inspecting the stellar quantities as discussed in the
\S 3. In particular, the V band magnitude is used to obtain estimates of the
distances to each star. The equivalent widths used by each analysis are compiled
for the elements and lines of each element, if available in the form of
electronic data tables on the web. These could be used for
re-analysis, if desired. The binary periods, as determined from the observed
variations in radial velocities between observations obtained at different
times, are compiled, although at present they are limited 
to a small number of systems.  We plan to pay special attention to reported 
periods in future updates of the SAGA database, as information on the role
that binary mass transfer may play on the observed surface abundances is thought
to be a critical factor in the identification of the decendants of the first
stars
\citep{Suda2004, Komiya2007}.

\begin{table}
  \caption{Physical quantities compile in the SAGA database}\label{tab:quantity}
  \begin{center}
    \begin{tabular}{*{3}{l}}
      \hline
	  Data table category & Item & Note \\
      \hline
      Bibliography & Title & \\
				   & Authors & \\
	               & Reference  & \\
      \hline
      Observing log & Object name & \\
                    & Observing date & Including Julian date and universal time \\
				    & Telescope & \\
				    & Resolution & \\
				    & Typical S/N & \\
				    & Exposure & \\
				    & Radial velocity & \\
      \hline
      Position      & RAJ(2000) & Right ascension (J2000.0) \\
                    & DEJ(2000) & Declination (J2000.0) \\
                    & $l$ & Galactic longitude \\
                    & $b$ & Galactic lattitude \\
      \hline
      Abundances & [X/Fe] & Enhancement of element 'X' relative to iron \\
	             & [X/H]  & Mass fraction of element 'X' \\
				 & $\loge$ & $\loge_{X} \equiv \log (n_{X}) + 12$ where $n_{X}$ is the number abundance of element 'X' \\
      \hline
      Atmosphere & $\teff$  & Effective temperature \\
	             & $\log g$   & Surface gravity\\
				 & \feoh      & Metallicity \\
				 & $v_{turb}$ & Velocity of microturbulence \\
      \hline
      Photometry    & Magnitude & U, B, V, R, I, J, H, K \\
                    & Color index & (B-V), (U-B), (V-R), (J-H), (H-K), (J-K), (V-K), (V-I), (R-I), E(B-V) \\
      \hline
      Equivalent width & EW & Equivalent width \\
                       & $\loge$ & Line abundance \\
                       & $\log gf$ & $gf$ values \\
                       & $\chi$ & Excitation potential \\
      \hline
      Binarity         & Binarity & Yes or No (unknown) \\
                       & Period & \\
                       & Radial velocity & \\
      \hline
      Solar Abundance  & Reference & Adopted or assumed value(s) \\
      \hline
    \end{tabular}
  \end{center}
\end{table}

We wish to remark on our treatment of the compiled object names. The name of a
given star is one of the key identifiers for each record in the database, but it
is not always unique. A few objects have several different names, according to
the catalogues from which they were drawn. Users of the SAGA database are
cautioned that the data are compiled and stored separately according to the
object name referred to in the {\it original} papers. As a result, our data retrieval
sub-system, described below, recognizes them as different objects. This may 
present some difficulties
when we discuss the properties of such stars. In order to make the user
aware of this problem, we put comments on the existence of multiple names in the
database, as provided by the script that analyzes the SIMBAD database. As
pointed out by \citet{Beers1992}, who referred to some HK survey stars as
``multiple identifications'', there are also cases in which the same catalogue
duplicates a given object (due, for example, to overlapping regions of the sky
being surveyed). The members of spectroscopic binaries are noted by the symbols
``A'' and ``B'', based on the notation employed in the original papers, and the
data are compiled separately for both members; only three such cases of binary stars
with information for both members are present in the current database. For stars
for which the abundances are derived using two or more different sets of stellar
parameters, we assemble the data separately by the use of appended symbols such
as ``d'' for dwarfs and ``s'' for subgiants at the end of its name.
This is the case for a few stars in the present database, for example,
HE~1327-2326 and HE~1300+0157.

\subsection{The database sub-systems}

The database consists of four sub-systems: (1) the reference management
sub-system, (2) the data compilation sub-system, (3) the data registration
sub-system, and (4) the data retrieval sub-system. The first three sub-systems
are used to compile and verify the observational data from the literature, while
the last one is for the use of the database. These sub-systems are described
briefly below.

\subsubsection{Reference management sub-system}

The reference is listed, based on a search of the papers conducted with the
Astrophysics Data System (ADS) Abstract Services\footnote{http://adsabs.harvard.edu/abstract\_service.html}
which report on observations of
metal-poor stars. Then, we check the candidate papers to see whether they have
included the required information, such as the results of abundance analyses
conducted at sufficiently high spectral resolving power. At present, we have
endeavored to collect papers that report stars with $\feoh < -2.5$. We intend to
increase the upper abundance limit in future updates of SAGA. For the papers
that pass these requirements, a unique reference code is registered on the web
system. Most of the papers added to the list are assembled by the data
compilation sub-system, except when the required data are not available. The
data are collected for all stars that are contained in the papers, including
those stars with higher reported metallicities.  

\subsubsection{Data compilation sub-system}

The observational data are taken from each paper accepted by the reference
management sub-system, and stored as comma separated values (CSV) files. The
data compilation sub-system
enables a set of data editing routines to input data easily by copying data
electronically. Human data editors pick up the required data (listed in
Table~\ref{tab:quantity})
from the papers by using an interface appropriately tailored for the
targets of the compilation. For example, when data editors store the data from the
numerical data table, they use the interface to convert the data table into CSV
files. The data table converter can deal with many types of electronic data table
formats available online, which differ from paper to paper as
well as from journal to journal. 
We define each set of data files of
the abundances and stellar parameters for a given reference and object; 
the reference and object name are the primary keys of the database.

\subsubsection{Data registration sub-system}

The stored CSV data files are registered into the database server through
use of the data registration sub-system run by the SAGA database administrator.
We have adopted the publicly available relational database management system
{\it MySQL}. During the registration process we convert the stored data into
several additional useful outputs for the data retrieval sub-system described
below. For example, the abundance data, given in units of [X/Fe], \loge,
or [X/H] from each paper, are converted into each other with the use of solar
abundance data. For now, we have adopted the solar abundances of
\citet{Grevesse1996}.
For example, the quantity [X/H] (where X corresponds to the species under
consideration) is computed from \loge\ automatically in the registration script
as [X/H] = \loge (X) - \loge (X,$\sun$).

At the same time, HTML files that include the original
information on the objects in the literature, and summary figures of the
abundance distribution for individual elements, are generated for each object.
These files can be used to obtain a quick review of the data included in each
paper, or for each object, so that users can trace the links to the data files
and abundance distributions for stars in the database, and easily access the
information on the paper(s) from which they were drawn.

\subsubsection{Data retrieval sub-system}

\begin{figure}
  \begin{center}
    \FigureFile(0.4\textwidth,){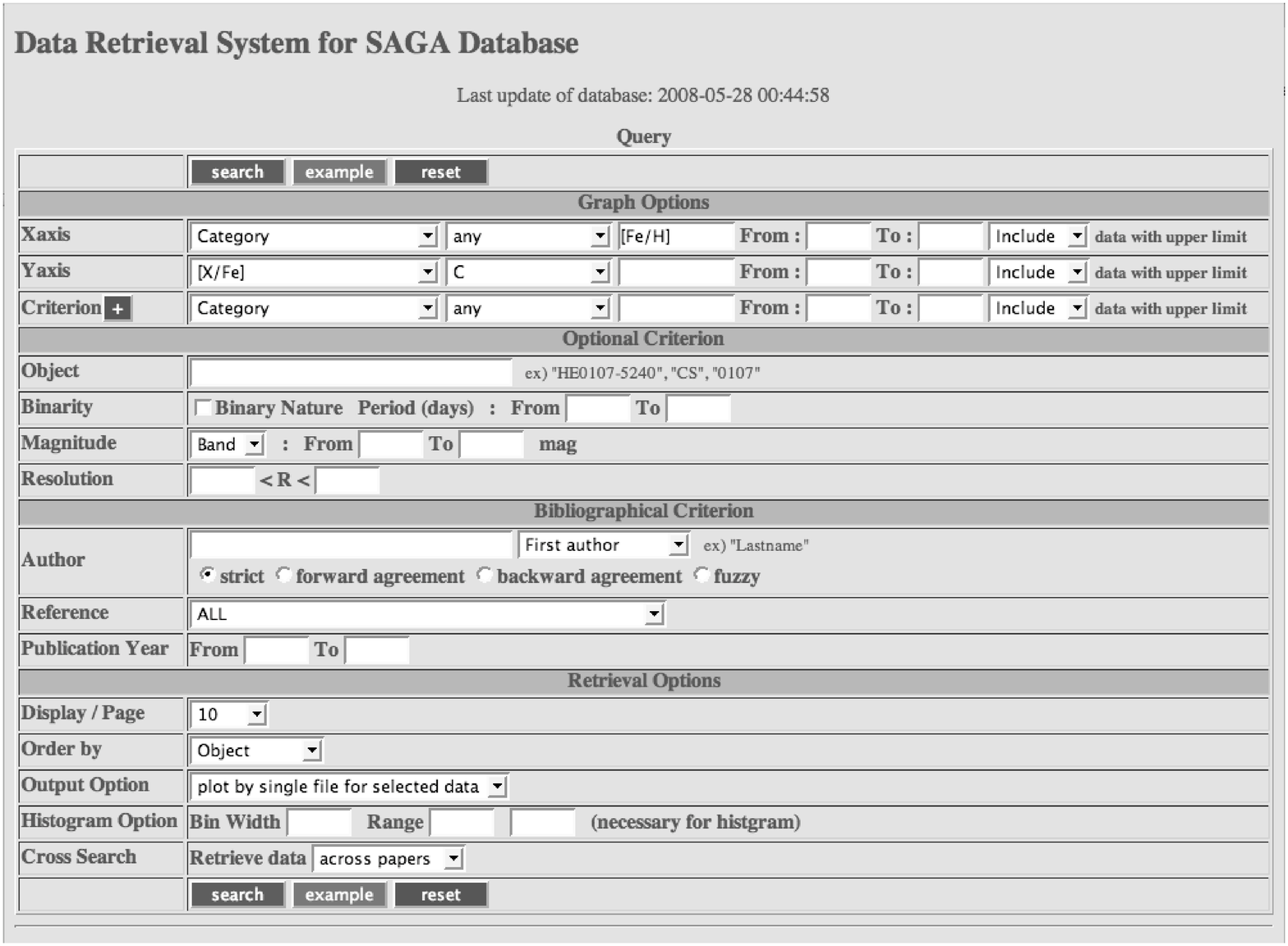}
  \end{center}
  \caption{Screen snapshot of the top page of the data retrieval sub-system 
for the SAGA database.
  }\label{fig:top}
\end{figure}

We constructed a web-based data retrieval sub-system for EMP stars based on a
script written in {\it Perl/CGI}. Users can access and select data based on
various criteria, and then inspect the selected data on a diagram with
user-specified axes. A screen snapshot of the query form is shown in
Figure~\ref{fig:top}. The first section of the form is used to specify the
quantities to draw in the graph, the axes in the first three lines, and the
criteria for object selection in the additional lines. One can choose any
desired quantity among those listed in Table~\ref{tab:quantity} to plot, e.g.,
the abundances of any elements (in any units of [X/Fe], [X/H], and $\loge$), the
atmospheric parameters, the photometric bands, the binary periods, and also the
stellar position and distance from the Sun estimated from the observed data. In
the first column, various categories of data can be chosen as axes. In the
second and third columns, one can specify the quantity to be plotted. In the
third column, one can specify the quantity directly by, for example, setting the
form input to ``[Fe/H]''. For elemental abundance data, it is also possible to
define new quantities beyond those listed in the original papers. For example,
users can obtain abundance ratios, such as [Ba/Eu] or [(C+N)/Fe], or even
[(Pb+Ba)/C], whose values and errors are calculated internally from the
corresponding data. The 4th and 5th columns set the range of values for each
selected quantity, with the option in the 6th column of whether to include or
exclude the objects that have data with only an upper limit reported. In the 4th
line, users can specify the required range in the data, if necessary, to select
or remove the objects from plotting, e.g., by setting $+0.5
\leq [{\rm C}/{\rm Fe}] \leq +2.0$.  The number of
criteria can be extended to as many as desired by the user. In the second
section of the form, one can set any additional criteria desired,
such as the object name, binarity status
and reported period, photometric magnitude in any bands, and the resolving power of
the observational setups used. The third section specifies the bibliographic
criteria. Through the use of these retrieval options, users can extract the data
containing specific object name, author, and the range of the year of
publication. Retrieval options are set in the fourth section, such as the number of
data to display in the resulting list and order of the list. By selecting the
output option, it is possible to obtain the distribution of the quantities in
the form of a histogram with an arbitrary size of bin width and range.

Cross-matched retrieval and plots are also possible as an option, which allows
for the extraction of data from different papers. This might be used for
checking the relationships between parameters when the object name is common
between several papers, but its data are not. Users should bear in mind that
the reliability of the results may be degraded due to the difference in the
observational setups used by the individual observations. 

\begin{figure}
  \begin{center}
    \FigureFile(0.4\textwidth,){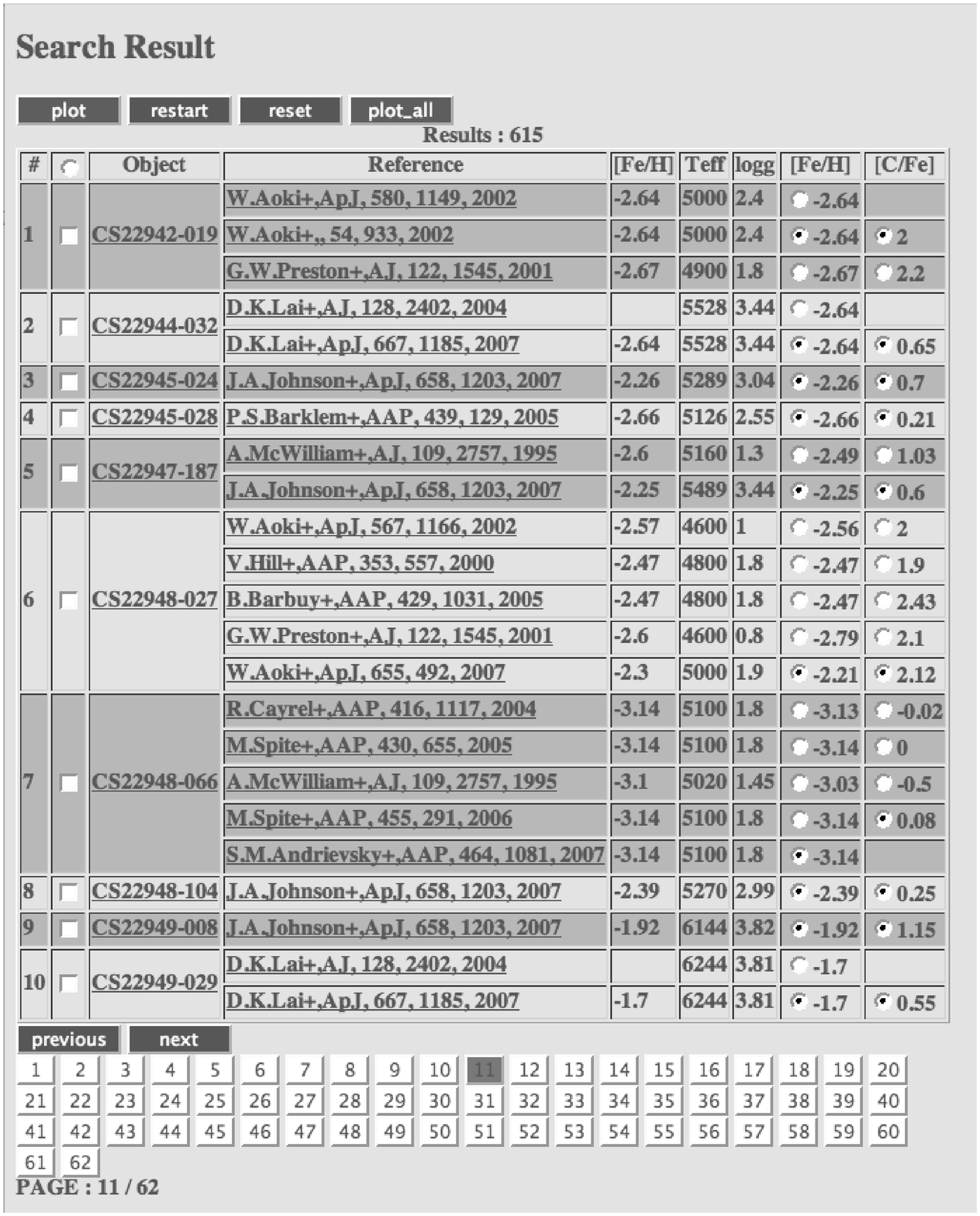}
  \end{center}
  \caption{Screen snapshot of the search result of the data retrieval sub-
system of the SAGA database.
In this case, the X-axis and Y-axis are set to [Fe/H] and [C/Fe], respectively.
  }\label{fig:result}
\end{figure}

A screen snapshot of an example retrieved set of records is shown in
Figure~\ref{fig:result}. The retrieved records from the {\it MySQL} server are
displayed in table format on the browser. The columns represent, from left to
right, the checkbox to select data to be plotted, the object name, the reference
code of the paper, the values of [Fe/H], $\teff$, and $\log g$, as well as the values
for the quantities selected as the axes of the plotted diagram.
The data to be plotted can be selected by the radio buttons that are
set initially based on the criterion such as the published year of derived data,
the adopted lines for element species, and the size of data errors.
By using the provided links to the object names and reference codes, one can trace the
information on object and reference, respectively.
For selected data, the
diagram is drawn in the web browser according to the choice of options, using
the publicly available graphic software {\it Gnuplot}. By specifying the
appropriate plotting option, users can plot the selected data sets from two or
more papers that match the selection criteria for each object. Graphs drawn in
the browser are equipped with simple functions for editing. The standard options
are to change the labels, the legends, and the scales and ranges of the graph.
The automatic links to the information on objects are generated in all
data points on the graph. Users
can also download the figures in various formats (png, eps, ps, and pdf, in 
color or in black and white), and can download the data from the form to their
local computer. Plotted numerical data, as well as the script to reproduce the figure
plotted are available from the server, if one wishes to edit the graph in more
detail. Numerical data are also accessible by tracing the link to each data set
in the list.

It should be noted that we temporarily use the single data set for given
element from one paper in the data retrieval system, and preferably adopt 1D LTE abundance.
Therefore, we adopt only one element abundance for single object for
plural element abundances such as different ionization status, while we
compile all the data for any kinds of ioniazation status and molecular lines.
In the current system, we prefer to adopt lower ionization state, i.e., [Fe I/H]
abundance is adopted in plotting the viewgraph if [Fe I/H] and [Fe II/H] abundances
are available in the database.
For atomic and molecular lines, we currently adopt the C II, CN, and O I abundance
rather than CH, NH, and OH abundance for carbon, nitrogen, and oxygen abundance, respectively.
This may sometimes cause inconvenience but it will be removed in the future update
so that the users can choose one of the adopted abundances for selected paper and object.
Of course, users can check the element abundances with different lines
by the quick review files of the database stated above and by the original
papers linked from them.

\section{Global characteristics of known EMP stars studied at high spectral
resolving power}

In this section we discuss the basic properties of the EMP stars in the SAGA
database. Here we focus on the characteristics of the sample stars as a whole, and
defer detailed analyses of the elemental abundances of individual stars, and 
discussion of the insight gained from them, to subsequent papers. All of the
figures presented below are obtained by use of the data retrieval sub-system.
For the objects with multiple sources of data, the values plotted are adopted
from the most recent papers.    

\begin{figure}
  \begin{center}
    \FigureFile(90mm,){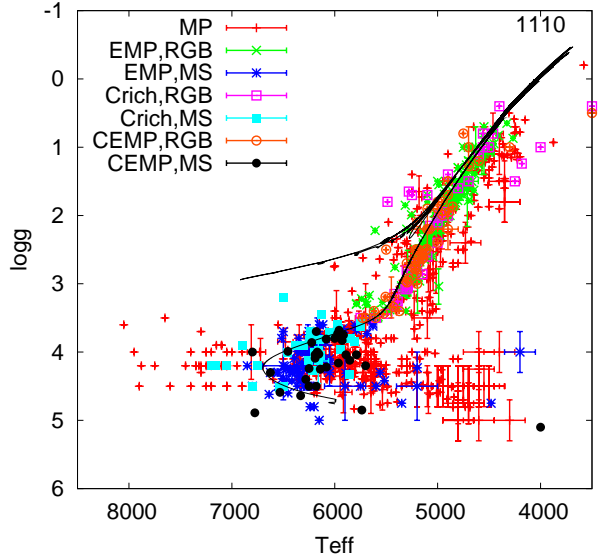}
  \end{center}
  \caption{Distribution of SAGA database stars in the $\teff$ vs. $\log g$ diagram,
           as plotted by the data retrieval sub-system.
           $\teff$ and $\log g$ are taken from the atmospheric parameters
	   adopted in the original papers.
	   A stellar evolutionary track for a model with 0.8 $\msun$ and $\feoh = -2.3$ is superposed
	   for comparison.
	   The number in the top right corner denotes the number of sample stars
           included in the plot.
	   The legend indicates our adopted classifications according to the
	   evolutionary status and abundance characteristics.
	   Details are provided in the text.
  }\label{fig:cmd}
\end{figure}

Figure~\ref{fig:cmd} shows the locations of the sample stars in an effective
temperature vs. surface gravity diagram. There are 1110 stars with these stellar
parameters measured from among the \nobj\ stars currently registered in the SAGA
database. In our system, the objects are classified according to 
evolutionary stage and their iron and carbon abundances.  For
convenience of the present discussion, we adopt the following conventions. The
objects are classified into dwarfs and giants (labeled ``MS'' and ``RGB'',
respectively), according to the requirement that giants have $\teff
\leq$ 6000 K and $\log g \leq$ 3.5.  We set the boundary between the EMP star
(labeled ``EMP'') and the other population II stars at $\feoh = -2.5$, which is
based on stellar evolution models for low- and intermediate-mass in which 
proton mixing driven by helium burning occurs for \feoh $\lesssim -2.5$
\citep{Fujimoto2000,Suda2004}.  
We also consider those stars with $\cfe \geq +0.5$ as ``CEMP'' and ``C-rich''
according to the metallicity $\feoh \le -2.5$ and $>-2.5$, respectively.
It should be noted that the criterion of carbon enrichment is different from that in
\citet{Beers2005} who defined $\cfe \geq 1.0$.
Accordingly, we have seven classes of objects, i.e., ``CEMP RGB'', ``CEMP MS'',
``EMP RGB'', ``EMP MS'', ``C-rich RGB'', ``C-rich MS'', and ''MP''; the last
label ``MP'' denotes the other metal-poor stars that are neither classified as
``EMP'' nor ``C-rich'', irrespective of their status as ``RGB'' or ``MS''.

Among the sample stars, EMP stars fall mostly in the range of $4500 < \teff <
6500$ K, which is significantly narrower than the range covered by the more
metal-rich stars. This is due to observational bias in the selection of targets
for high-resolution spectroscopy, as researchers have naturally favored studies
of the most extreme stars in the past decade. In Fig.~\ref{fig:cmd}, theoretical
evolutionary tracks are plotted for a model star of $0.8 \msun$ and $\feoh =
-2.3$. The EMP stars occupy the evolutionary track near the main-sequence
turnoff and on the lower giant branch. The dearth of sample stars below $T
\simeq 4500$ K, including near the tip of the RGB, is related to the difficulty in
obtaining abundance information for many elements due to the presence of strong
molecular features. One exception is the dwarf G~77-61, with $\teff = 4000$ K,
according to \citet{Plez2005}, who revised the metallicity from the value $\feoh = -5.6$ assigned by
\citet{Gass1988} to $\feoh = -4.01$.  Stars on the AGB are also scarce among EMP
stars, other than CS~30322-025
\citep[with $\feoh=-3.5$ and $\teff = 4100$ K]{Masseron2006}. 
At higher temperatures, our sample of EMP stars lacks blue
stragglers beyond the turnoff, while
they are abundant among the more metal-rich sample stars. In addition, the
horizontal-branch stars with temperatures $\teff \gtrsim 5500$K are absent in the
Fig.~\ref{fig:cmd}, again attributable to the selection bias associated with follow-up
observations. Such stars have not been studied in great detail at high
resolution, even though targets exist, e.g., from 
\citet{Beers1992}. At low metallicity, stars of higher temperature do not display
a wide range of detectable absorption features associated with heavier elemental
species, so they are understandably not preferentially chosen for high-resolution
follow-up.

\begin{figure}
  \begin{center}
    \FigureFile(90mm,){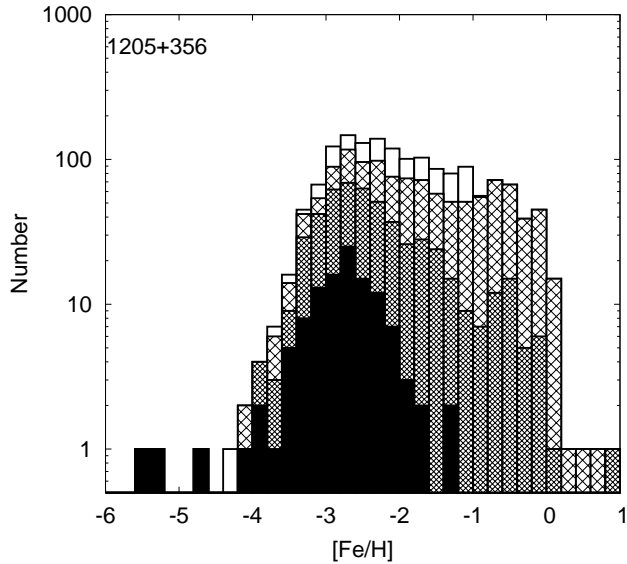}
  \end{center}
  \caption{The metallicity distribution ($\feoh$) for the stars
           currently registered in the SAGA database.  A total
	   of 1205 stars with measured iron abundances,
           denoted by hatched boxes,
	   are divided into three groups -- C-rich stars
	   defined by $[{\rm C} / {\rm Fe}] > +0.5$ (filled boxes),
	   giants defined by $\teff < 6000$ K and $\log g < 3.5$
	   (shaded boxes), and an 
	   additional 347 stars from the \citep{Frebel2006a} bright metal-poor sample
	   and the C-rich sample in \citet{Lucatello2006} (open boxes).
	   Group classifications are not taken into 
           account for these latter samples.
  }\label{fig:histFe}
\end{figure}

Figure~\ref{fig:histFe} shows the metallicity distribution for the 1205 sample
stars\footnote{The present database does not include iron abundance
for 7 stars because it is not given in the original paper.} 
with available iron abundance measurements. The distribution peaks near
$\feoh \simeq -2.7$; the decrease of sample stars at higher metallicity is an
artifact of selection effects. Since a confident assignment of 
metallicity for stars with $\feoh < -3$ can only
be determined from high-resolution spectroscopy, a general tendency exists
for a decreasing number of metal-poor stars at lower metallicity.
The sharp drop discernible for $\feoh < -4$ 
reflects the metallicity distribution function of field
halo stars identified to date; only 5 stars are found below
this metallicity 
(including HE~0557-4840, with $\feoh = -4.8$, discovered by \citet{Norris2007},
and two stars very close to the boundary, CD-38$^{\circ}$245 with $\feoh = -4.2$, and
G~77-61 with $\feoh = -4.03$), while $\simeq 150$ stars are known with metallicities in
the range $-4 <
\feoh < -3$. In this Figure, we have included the medium-resolution
follow-up observations of ``saturated stars'' ($9 < B < 14$) from the HES
provided by \citet{Frebel2006a}, and the cooler, carbon-rich stars from the HERES
data sample of
\citet{Lucatello2006}.
Inclusion of these data, denoted by open boxes, increases the apparent steepness
of the metallicity distribution function at $\feoh \gtrsim -4$. Note that the
data from these authors lack some of the information on stellar parameters;
these samples are excluded in the following discussion.

\begin{figure}
  \begin{center}
    \FigureFile(90mm,){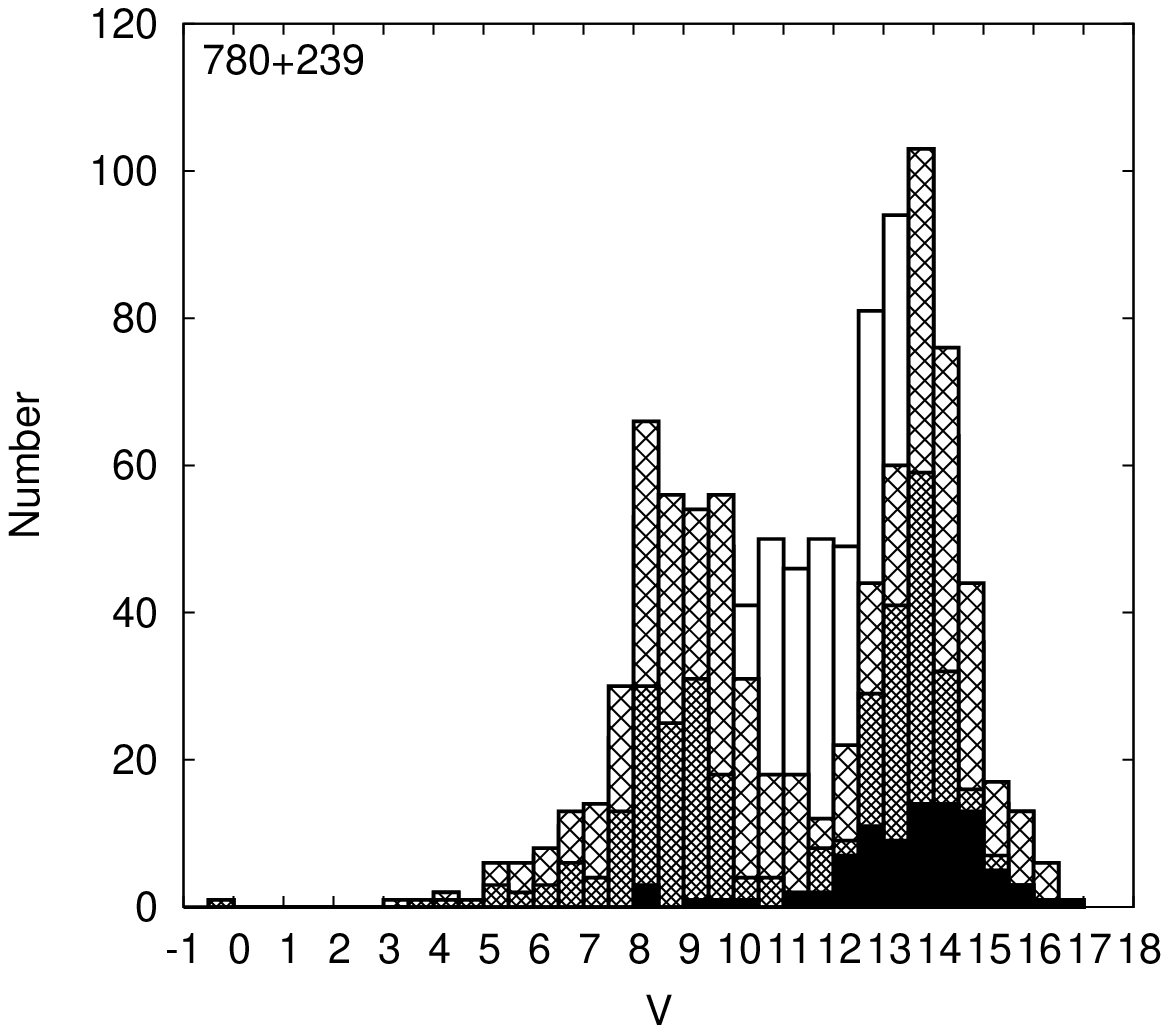}
  \end{center}
  \caption{Distribution of the V magnitudes for stars in the current SAGA
database.  There are 866 stars in our sample with known V magnitudes. The
                   meanings of the boxes are the same as in Fig.~\ref{fig:histFe},
				   but the stars from \citet{Lucatello2006} are not
                   included. The brightest star is HD~124897, which has $V =
                   -0.05$. For the 239 bright metal-poor sample of
                   \citet{Frebel2006a}, V band magnitudes are computed from the B
                   magnitudes and (B-V) colors available in the literature.
                   }\label{fig:histV}
\end{figure}

In Figure~\ref{fig:histV} we show the V magnitude distribution of the 866 stars
included in the SAGA database. It exhibits a bimodal distribution, with two
peaks at $V \simeq 8-9$ and $13-14$, which reflects the history of past survey
efforts. Stars in the brighter peak are primarily those discovered by
\citet{Bond1980}, with limiting magnitude $B \simeq 10.5$, and from the
high-velocity star survey
\citep{Carney1981} and \citet{Carney1994}.
The deficiency of stars in the spectroscopic sample with apparent magnitudes
in the range $9 \lesssim V \lesssim 13$ 
is due to observational bias in later objective-prism
surveys. The HK survey and the HES are designed to search in the ranges of $12 <
B < 15.5$ \citep{Beers1985} and $14 < B < 17.5$ \citep{Christlieb2001a}. The apparent
dominance of dwarfs in between the two peaks can be explained by the sampling of
high-velocity stars in this range, since stars with high proper motions are
found almost exclusively among dwarfs. For the bright ($B < 14)$ stars on the
HES plates, the data recovery is now ongoing \citep{Frebel2006a}. The 286 stars
observed with medium- resolution follow-up spectroscopy to date partially fill
the gap around $V \sim 11$.

In the metallicity range $\feoh < -2.5$, our sample consists of a total of 369
EMP stars, which contains 253 giants and 116 dwarfs. Among them, there are 81 C-rich stars
(51 giants + 30 dwarfs). 
\citet{Komiya2007} theoretically approximated
the fraction of dwarfs (and subgiants, using the definition of $\log g \geq
3.5$), relative to giants, to be $\sim 40 \%$ for a limiting magnitude of 15
mag. This is in contrast to the flux-limited sample, with limiting magnitude of
15.5 mag from \citet{Beers1992}, whose data show that approximately two-thirds of
their sample are turn-off stars (due to a known temperature-related selection bias).
In the SAGA database, among the stars with $V \leq 15.5$, dwarfs represent $54
\%$ of the total.

\begin{figure}
  \begin{center}
    \FigureFile(90mm,){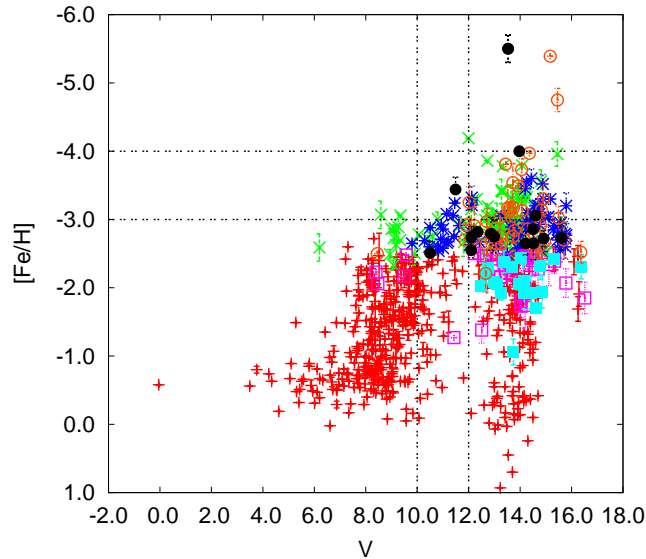}
  \end{center}
  \caption{Relationship between metallicity and V magnitude in 767 sample stars.
           The meanings of the symbols are the same as in Fig.~\ref{fig:cmd}.
  }\label{fig:FeV}
\end{figure}

Previous work (e.g., \cite{Lucatello2006}) has claimed that CEMP stars
occupy $\sim 20 \%$ of EMP stars. Note that the fraction of C-rich stars can be
as large as $\sim 30 \%$, if we take the number of stars with carbon detection as
a denominator, instead of the total number of EMP stars.
Note that many previous workers have adopted a more conservative
definition ([C/Fe] $> +1.0$) for the identification of CEMP stars than we
use in our present discussion ([C/Fe] $> +0.5$).
Figure~\ref{fig:histV} shows the large fraction of C-rich
stars in the SAGA datbase for $V \gtrsim 12$. Even though this is surely
influenced by a variety of selection biases, 
the discoveries of
the so-called ultra ([Fe/H] $< -4.0$) and hyper ([Fe/H] $< -5.0$) metal-poor
stars, all of which exhibit large carbon enhancements, suggests that the fraction of
CEMP stars is quite high at the lowest metallicities. 
It should be stressed that the observed fraction is unlikely
to be biased at metallicities
$\feoh \lesssim -3.0$, because of the difficulty in predicting abundances
with medium-resolution spectroscopy in this regime.  In any case, our present sample
confirms that the fraction of carbon-rich stars is larger at lower
metallicity.
In the current SAGA database, for $V \lesssim 10$, 
there are few stars with $\feoh \sim -3$, and no stars with $\feoh < -3.5$
(see Figure~\ref{fig:FeV}).

According to our selection criteria for assembling EMP stars from published
papers, the total fraction of carbon-rich stars ($\cfe \ge +0.5$) is
13.4 \% (24.1 \% among stars with derived carbon abundance).
Among them, C-rich EMP stars
($\feoh \le -2.5$) comprise 22.0 \% (28.6 \%) of the total
sample of EMP stars. In particular, C-rich EMP giants occupy
20.2 \% (21.4\%) of EMP stars, while C-rich EMP dwarfs
represent 25.9 \% (66.7\%). The extreme discrepancy of
C-rich fraction among the C-detected sub-sample is also found for the entire sample
(20.4\% for giants and 33.7\% for dwarfs), and is due to the sensitive
dependence
of the detection limit on [C/H] for increasing effective temperatures
(see Fig.~11 of \cite{Aoki2007b}).
Even if we change
the criteria on metallicity and carbon enhancement to $\feoh
\le -3.0$ and $\cfe
\ge +1.0$, a similar trend is obtained (16.8 \% (21.4\%) in total,
16.3 \% (16.9\%) for giants, and 18.2 \% (66.7\%) for dwarfs),
although the sample size becomes small (125 (98) stars with
$\feoh \le -3.0$). 
These two fractions for different criteria may possibly give the
lower and upper limit for the true C-rich fraction.
These values may be ascribed to the motivations of observers, i.e.,
observers may have specifically intended to detect carbon in order to investigate
the characteristics of C-enhanced stars. Accordingly, the C-rich
fraction among the total sample has been increased by re-investigations of the
known C-enhanced sample.
On the other hand, C-rich features can be seen in the initial 
low- to intermediate-resolution spectra, selected from, for example,
\citet{Christlieb2001b},
which may argue that the current C-rich fraction among C-detected sample 
in the SAGA database may be a reasonable approximation of reality.

\begin{figure}
  \begin{center}
    \FigureFile(90mm,){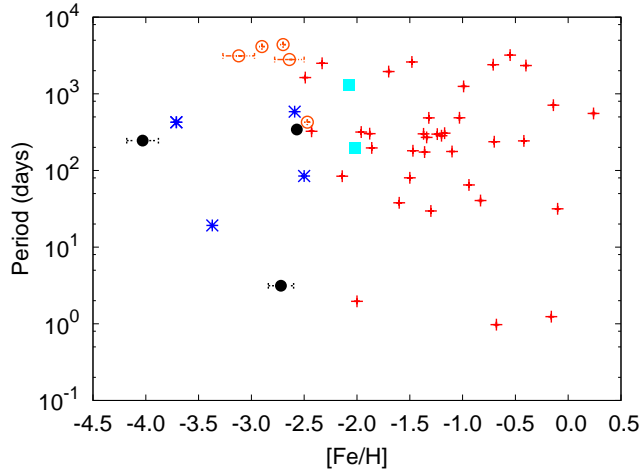}
  \end{center}
  \caption{Binary period distribution of for the 51 stars with measured periods
in the SAGA database.
           The meanings of the symbols are the same as in Fig.~\ref{fig:cmd}.
  }\label{fig:binary}
\end{figure}

\begin{figure}
  \begin{center}
    \FigureFile(150mm,){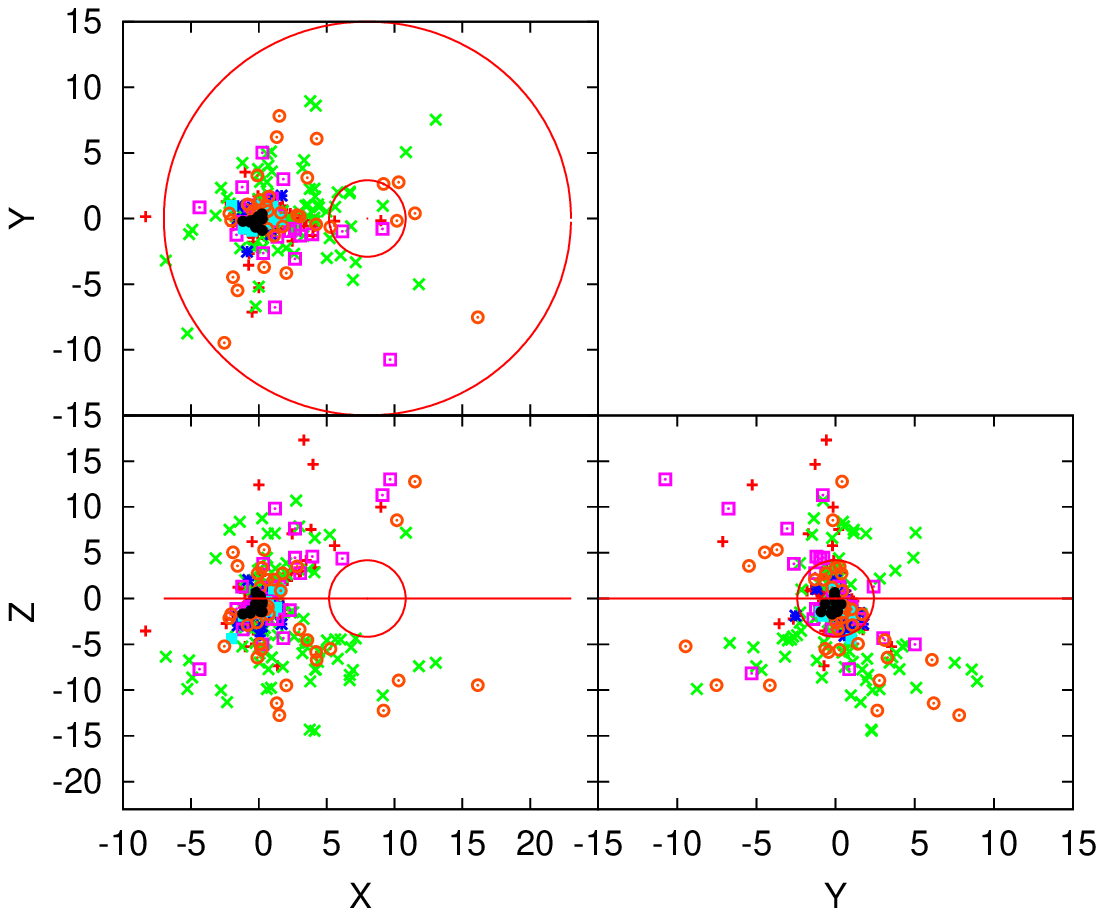}
	\vspace{10mm}
  \end{center}
  \caption{The spatial distribution of 708 stars from the SAGA database. Distances to stars are estimated by
 		   assuming that all stars have 0.8 \msun.
		   The coordinates X, Y, Z are set at the local standard of rest and
		   are directed to the Galactic center, counter-rotation direction, and
		   north Galactic pole, respectively.
  		   The Galactic disk and center are also schematically shown by lines and circles.
           The meanings of symbols are the same as in Fig.~\ref{fig:cmd}.
  }\label{fig:position}
\end{figure}

There are 51 known binaries present in the SAGA database. This is a
surprisingly small value (less than 10\% of the total sample), when we consider the
large binary fraction ($\sim$ 50\%) among nearby stars of younger populations.
One of the primary reasons is the small number of EMP stars that
have received attention from velocity monitoring programs in the past
decade (and the short span in time of monitoring programs for those that have).
Another reason could be that EMP stars may include a large fraction
of long period binaries, and the resulting difficulty in detection of
changes in their observed radial velocities. 
It has been
suggested by \citet{Komiya2007}, from the standpoint of stellar evolution, that most
EMP stars were born as the low-mass members of binary systems, and only those with
large orbital separations could have survived the mass-transfer event when the
primary stars swelled at the end of their lives. Observationally, it has also
been reported by
\citet{Preston2000} that C-rich stars among the sample of blue metal-poor 
stars exhibit longer periods than those of their C-normal counterparts. Binary
periods have been estimated for 51 stars in the SAGA database, and range from
$\sim$ 1 to $\sim$ 4000 days.  The distribution of 
periods is shown in Figure~\ref{fig:binary} as a function
of metallicity. Although the sample is very small, it clearly
shows a difference in the observed binary periods between giants and dwarfs,
with giants having systematically longer periods than dwarfs. Interestingly,
there are 4 CEMP giants with long periods, in excess of $P > 2500$ days, 
while 7 EMP and CEMP
dwarfs occupy the region with $P < 1000$ days. The lack of short-period binaries for
giants could be understood in terms of their larger radii
by supposing that the binary separations must be larger than the radius of
the giants.
On the other hand, the
lack of long-period binaries for dwarfs may possibly be attributed to
the length of time they have been monitored, which should be improved in 
the future. For stars
with $\feoh > -2.5$, some blue metal-poor stars show long periods of $> 1000$
days. Since, as argued in \citet{Komiya2007}, binaries may have played an
important role in the early epochs of the Galaxy, much more data on the binarity
and period distributions are desired.

The spatial distribution for 708 stars from the SAGA database is shown in
Figure~\ref{fig:position}. The distance is computed from the V band and the
stellar luminosity, the latter of which is derived from the effective
temperature and the surface gravity of the model atmosphere by assuming a
stellar mass of $0.8 \msun$. The coordinates are taken from the literature or
from the SIMBAD database. In some cases, the Galactic coordinates are computed
from the equatorial coordinates. The plotted sample includes 203 EMP
stars. Among them, only 36 and 14 stars are CEMP giants and CEMP dwarfs,
respectively. The maximum distance to the dwarfs is estimated to be $\approx$ 5
kpc from the Sun, while the maximum distance is $\approx$ 28 kpc for giants.

\section{Summary and Discussion}

We have constructed the SAGA (Stellar Abundances for Galactic Archeology)
database of extremely metal-poor stars in our Galaxy. The compiled data are accessible on
the web and are opened to all researchers now. Our database
includes information on observational details, abundances, atmospheric
parameters, photometry, equivalent widths, and binarity status and periods.
These data are taken from published papers, with the use of a web-based system
of data compilation equipped with useful tools to convert them from various
forms of electronic data tables into CSV format. A data retrieval system has
been developed which enables the retrieval and plotting of the data selected
according to various criteria.

Our sample includes \nobj\ stars with distinct object names, roughly half of which
are giants. The number of giants becomes twice as large as that of
dwarfs if we consider only stars with $\feoh < -2.5$. The fraction of
carbon-enhanced stars ([C/Fe] $\geq$ +0.5) amounts to $\sim 30\%$ among the
sample of stars with derived carbon abundance for $\feoh < -2.5$. The sample
stars exhibit a bimodal distribution of V band magnitudes, which is ascribed to
the different coverage of effective magnitude range among the large-scale
surveys of metal-poor stars. There may exist different distributions of binary
periods among the stars with this information available. It is shown for stars
with $\feoh \lesssim -2.5$ that the binaries with a giant member have typically longer
periods than those with a dwarf member, and that there are no dwarfs in binaries having
periods of $> 1000$ days yet confirmed. 
Considering the spatial distribution, our
sample may have some biases for the discussion of the properties of metal-poor
stars because of the different sampling volumes for dwarfs and giants. In fact,
we only have detailed elemental abundances for dwarfs within $\lesssim$ 5 kpc
from the Sun, while giants cover distances extending to more than $\gtrsim$ 20
kpc in the current sample.

Since the EMP stars in our Galaxy are useful probes for our understanding of the
chemical and formation history of our Galaxy, large increases in such
data are desired by observers and theoreticians alike. A number of observing
projects are planned to increase the stellar sample. For example, the stellar
extension program of the Sloan Digital Sky Survey, SDSS/SEGUE, is obtaining
medium-resolution spectra from which additional EMP stars may be selected to a
depth of up to $\sim$ 100 kpc \citep{Beers2004}. LAMOST \citep{Zhao2006} is a multi-fiber 4m
telescope project that will enable up to 4000 stellar spectra to be obtained
simultaneously in each exposure. The total survey effort is planned to encompass
several million stars. These projects will increase the number of candidate EMP
stars by several orders of magnitude in the near future, as compared with the
number of known EMP stars known at present. High-resolution spectroscopic follow-up
will be required, making use of dedicated programs, such as the proposed WFMOS
effort on the Subaru telescope, and with the next generations of Extremely Large
Telescopes, with diameters of 30m or more. 

\begin{figure}
  \begin{center}
    \FigureFile(\textwidth,){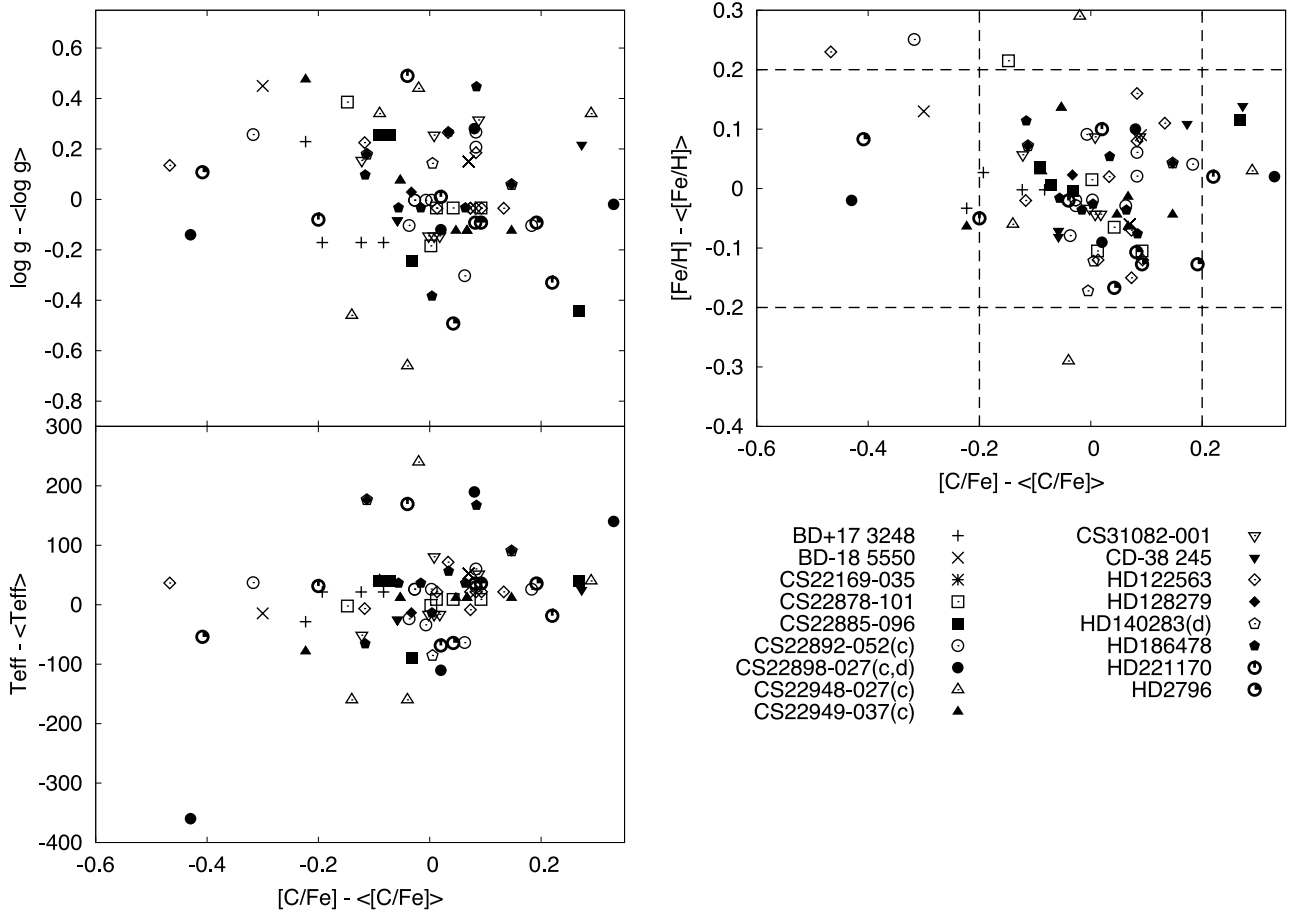}
  \end{center}
  \caption{Consistency check of the independent analyses for selected 17 stars
  from SAGA database. Each value denotes the deviation from the average value
  of adopted or derived quantity in each object for which it is reported.
  The symbol "c" and "d" in the legend of object names denotes the ``carbon-rich''
  and ``dwarf'', respectively.
  Othere stars without symbols are giants without carbon enhancement.
  Note that some of the papers adopt the same set of atmospheric parameters
  and abundances. Such data are completely overlap with each other in this figure.
  Note also that the data without values reported by authors do not apper in the
  figure, which is sometimes the case for $\cfe$.
  The majority of the data points are located within the typical errors as enclosed
  by auxiliary dashed lines in top right panel.
  }\label{fig:analyses}
\end{figure}

It is important for us to understand how large the discrepancies are caused
by the independent analyses.
In order to check the difference among the derived abundances by different authors, 
we pick up 17 stars from our sample and compare their derived carbon abundances.
The 13 stars among them are retrieved from the sample for which more than 8 papers
report the abundance analysis.
The 12 stars of them are giants and do not show carbon enhancement.
The remaining 4 stars are added to cover the various EMP stars.
They are carbon-enhanced stars of giant and dwarf (CS 22948-027 and CS 22898-027, respectively),
an extremely iron-poor star having $\feoh \sim -4$ (CD-38$^{\circ}$245),
and normal giants for which carbon abundance is reported by more than 5 papers (CS 22169-035).
In Figure~\ref{fig:analyses}, we show the deviations from average values of $\log g$,
$\teff$, and $\feoh$ as a function of those of $\cfe$.
It should be noted that all reported carbon abundance is based on 1D LTE model atmosphere
and most of the analyses adopt the synthetic spectral technique using CH G band.
The majority of the plotted stars are located within the 0.2 dex for $\cfe$ value,
which is well explained in terms of the errors associated with the different
values of atmospheric parameters and of the usage of different solar abundance
from paper to paper, the latter of which can be important for CNO abundance.
Some of the large deviations of the adopted or derived values can also be
explained by the analyses based on low-resolution spectra
(for example, the case of CS 22898-027),
although, the reasons for different results are not necessary obvious for all cases.
As can be seen in the left panels of Fig.~\ref{fig:analyses}, the large differences
are highly correlated with the deviations of the adopted values of atmospheric parameters,
the latter of which is due to the different way of analyses and corrections for
the determinations of surface gravity and effective temperature.
In fact, for CS22948-027, the largest discrepancy of atmospheric parameters and
metallicity in the figure seems to be caused by both the different method and
correction of them.

Accordingly, users should be warned about the possible discrepancy of independent
analyses when they use the combined data derived by different authors.
At any situations in using our database, users can go back to the original papers
at the data retrieval system and check the information on analyses and discussion.
The extreme case of discrepancies, if happens, will be discussed in the latest
original paper by comparing with the previous works.
Therefore, we will not continue to discuss here about the systematic differences between
previous works for all objects and elements in the database.
For the abundance deternimations with non-LTE scheme or with 3D model atmospheres,
their effect for extremely metal-poor stars should also be mentioned in considering
the different analyses in more detail.
However, it is beyond the scope of this paper and is discussed in the other
extensive works for non-LTE abundances (see, e.g, \cite{Andrievsky2007}) and for 3D model
atmosphere \citep{Asplund2001,Collet2007}.

At present, we are planning to include the information on the analyses adopted
by the authors and to implement the option of choosing the LTE or NLTE abundances.
On the other hand, It is desirable to improve the quality of data by creating
a homogenized dataset that enables us to refine the statistical analysis of abundance trends.
For users of interest, we can provide the compiled data related to equivalent width
and other necessary data for their re-analysis of the sample.

We plan to continually update the SAGA database with updates as new papers
reporting on high-resolution spectroscopic follow-up appear in the literature.
We also plan to continue an effort to provide more complete coverage of existing
data, by supplementing the SAGA database with stars of higher metallicity, and
by extending the temporal coverage to circa 1990. In forthcoming papers, we plan
to use the updated SAGA database to discuss more thoroughly the abundance trends of EMP
stars, and compare them with 
theoretical models.\\

We are grateful to T. C. Beers for reading the manuscript and for giving
helpful suggestions and comments including the denomination of the database.
We thank S. Lucatello for kindly providing the abundance data of
carbon, nitrogen, and iron in the Hamburg/ESO R-process Enhanced Star
(HERES) survey sample.
We are also grateful to the anonymous referee for his/her suggestion about
the influences of independent analyses.
This research has made use of the ADS
database, operated at SAO/NASA, USA, mirrored by NAOJ, Japan, and SIMBAD and
VizieR database, operated at CDS, France. This work has also made use of the
observations with low- to high-dispersion spectroscopy by the optical telescopes
all over the world. This work has been partially supported by Grant-in-Aid for
Scientific Research (15204010, 18104003, 19740098), from Japan Society of the
Promotion of Science.

\bibliographystyle{apj}
\bibliography{apj-jour,reference}

\end{document}